\documentclass[amsmath,amssymb,
onecolumn,apm]{revtex4-2}

\usepackage{graphicx}
\usepackage{caption}
\usepackage{xcolor}

\begin{document}

\title{Matter--Geometry interplay in new scalar tensor theories of gravity}

\author{Mihai Marciu}
\email{mihai.marciu@drd.unibuc.ro}
\affiliation{ 
Faculty of Physics, University of Bucharest, Bucharest-Magurele, Romania
}

\date{\today}

\begin{abstract}
The paper studies the possible interplay between matter and geometry in scalar tensor theories of gravitation where the energy--momentum tensor is directly coupled with the Einstein tensor. After obtaining the scalar tensor representation of the $f(R, G_{\mu\nu}T^{\mu\nu})$ gravity, the analysis continue with an approach based on the thermodynamics of irreversible processes in open systems. To this regard, various thermodynamic properties are directly obtained in this manner, like the matter creation (annihilation) rate and the corresponding creation (annihilation) pressure. In the case of the Roberson--Walker metric several analytic and numerical solutions are found in the asymptotic regime. In the last part of the manuscript a specific parametrization for the Hubble rate is constrained using the Markov Chain Monte Carlo algorithms in the case of cosmic chronometers (CC) and BAO observations, obtaining an approximate numerical solution which can describe the cosmological model. For this model, we have obtained by fine--tuning some numerical solutions which exhibit creation mechanisms in different specific regimes. 

\end{abstract}

\maketitle

\section{Introduction}
\label{intro}

\par 
In the present days the cosmological theories have reached an important point, affecting the development of science and technology \cite{ParticleDataGroup:2022pth, Bamba:2012cp, Berti:2015itd}. An important framework is represented by General Relativity \cite{Weinberg:1988cp}, a hundred years old theory which can explain various physical effects on local and galactic scales \cite{Will:2014kxa}. Although this model can successfully describe various physical characteristics on small and large scales, it has its own limits \cite{Carroll:2003wy}. In the cosmological theories the dark matter paradigm \cite{Navarro:1995iw, Bertone:2004pz, WMAP:2012nax} appeared in the late 70s, after several astrophysical observations pointed out that baryonic matter alone cannot explain the dynamics of the Universe at galactic levels \cite{EINASTO1974, Ostriker1974}. Since then, many studies have analyzed the possible candidates, opening a research area in this direction \cite{Clowe:2006eq, BOSS:2016wmc, Moore:1999nt, Klypin:1999uc, XENON:2018voc, Marciu:2016ige}. The accelerated expansion of the Universe \cite{WMAP:2010qai, Copeland:2006wr, Peebles:2002gy} is another curious phenomenon in modern cosmology, driving the large scale matter dynamics \cite{Padmanabhan:2002ji}. Discovered at the end of the last millennium, the dark energy phenomenon has been probed by various astrophysical observations \cite{Planck:2018vyg, SupernovaSearchTeam:2004lze, Lewis:2002ah, Beutler:2011hx} and theories \cite{Nojiri:2010wj, Nojiri:2006ri}. 
\par 
The most simple model is represented by the $\Lambda$CDM model \cite{Copeland:2006wr}, where the Einstein--Hilbert action is complemented by a cosmological constant term \cite{SupernovaSearchTeam:1998fmf}, driving the acceleration of the Universe. Although it is a promising theory, it has its own drawbacks \cite{Peebles:2002gy}, due to the constant equation of state for the dark energy sector \cite{Copeland:2006wr}. Most astrophysical observations \cite{SupernovaSearchTeam:2004lze} have pointed out that the dark energy sector is described by a dynamical equation of state \cite{Linder:2002et}, and the $\Lambda$CDM model can be regarded only as an effective approximate theory. Taking into account the dynamical evolution of the dark energy equation of state \cite{Frieman:2008sn}, from an astrophysical point of view the latter term can be associated to quintessence \cite{Li:2011sd}, phantom \cite{Caldwell:2003vq}, and quintom \cite{Cai:2009zp, Marciu:2016aqq} behaviors. Within these theories, the dark energy sector is modelled as a time dependent field(s) minimally or non--minimally coupled \cite{Marciu:2022wzh, Marciu:2020vve, Marciu:2020ski, Bahamonde:2020vfj, Marciu:2019cpb} which drives the expansion of the Universe. Another approach in the cosmological theories is based on modified gravity models \cite{Capozziello:2011et}, where various invariant terms are encapsulated in the specific action, leading to different physical effects \cite{Nojiri:2017ncd}. The most simple model is the $f(R)$ theory \cite{Sotiriou:2008rp}, based on the scalar curvature, extending the Einstein--Hilbert action in a non--trivial manner. Since then, many alternative theories have been proposed \cite{Bahamonde:2019gjk,Marciu:2020ysf, Marciu:2018oks, Marciu:2017sji}, based on specific invariant terms. Most of these theories are compatible to the dark energy phenomenon \cite{Clifton:2011jh}, explaining the late time large scale dynamics of the Universe at the background level.
\par 
Within these theories \cite{Abdalla:2022yfr}, some particular extensions have appeared, based on the interplay between matter and geometry \cite{CANTATA:2021asi}. In the case of $f(R,T)$ theories \cite{Harko:2011kv, Jamil:2011ptc}, a specific coupling between scalar curvature and the trace of energy momentum tensor have been proposed. In the latter framework different authors have investigated specific physical features \cite{Houndjo:2011tu, Sharif:2012zzd, Alvarenga:2013syu, Shabani:2014xvi, Yousaf:2016lls, Harko:2014pqa}, confirming the viability of the interplay between matter and geometry. Another extension is related with the $f(R, T^2)$ framework \cite{Cipriano:2024jng, Roshan:2016mbt}, based on the energy momentum squared gravity. The energy momentum squared gravity has been analyzed in various cosmological applications in the recent years \cite{Board:2017ign, Moraes:2017dbs, Nari:2018aqs, Bahamonde:2019urw, Marciu:2023muv}. A more fundamental interplay between matter and geometry have been proposed recently \cite{Asimakis:2022jel, Marciu:2023jvs}, by considering a direct coupling between the Einstein tensor and the energy--momentum tensor \cite{Odintsov:2013iba}, a theory capable of explaining the late time acceleration, highly compatible with recent astrophysical observations. In this theory \cite{Haghani:2013oma}, matter and geometry is considered to be on the same level from an abstract point of view, offering a physical balance of the latter concepts \cite{Sharif:2013kga}. Recently, different authors have investigated various properties of $f(R,T)$ \cite{Harko:2021bdi} and $f(R, T^2)$ models \cite{Cipriano:2023yhv}, by adopting a specific point of view based on the thermodynamic properties of open systems \cite{Prigogine:1989zz}. Within these cosmological systems, the standard continuity equation is not satisfied, due to the interplay between matter and specific physical properties of the geometrical manifold \cite{Ayuso:2014jda}.

\par 
In the present paper we shall consider a possible interplay between matter and geometry at the level of background dynamics, encapsulating a direct coupling between the Einstein tensor and the energy--momentum tensor \cite{Asimakis:2022jel} in the corresponding action. After we present the geometrical representation of the $f(R, G_{\mu\nu}T^{\mu\nu})$ gravity \cite{Marciu:2023jvs} we shall discuss its scalar tensor representation, by introducing two time depending scalar fields. Then, by relying on the thermodynamics of open systems \cite{Prigogine:1989zz, Harko:2021bdi, Pinto:2022tlu}, we deduce various physical features \cite{Cipriano:2023yhv} in specific asymptotic regimes, discussing the compatibility of the $f(R, G_{\mu\nu}T^{\mu\nu})$ theory \cite{Marciu:2023jvs} with current astrophysical observations. The main aim of the present paper is to explore the possibility of matter creation or annihilation in the Universe as a consequence of a direct interplay between matter evolution and space--time geometry.

\par 
The structure of the paper is as follows. In Sec.~\ref{actiune} we present the equations of the $f(R, G_{\mu\nu}T^{\mu\nu})$ gravity, obtaining the scalar tensor representation. Then, we proceed with a short introduction into the thermodynamic properties of open systems, presenting the relevant relations for the analysis in Sec.~\ref{termo}. In Sec.~\ref{analiza} we study the analytical and numerical properties of our model, in asymptotic regimes. We also discuss the case where the field equations are numerically solved in a direct manner, confronting the CC+BAO observations. Finally, in Sec.~\ref{conclusions} we outline the physical effects found in the analysis and present the last concluding remarks.

\section{The action, the field equations, and the scalar tensor representation}
\label{actiune}

\subsection{The geometric representation of $f(R, G_{\mu\nu}T^{\mu\nu})$ gravity}

\par 
In this section we shall present various theoretical aspects \cite{Marciu:2023jvs} related to our cosmological model. In what follows we shall take into account the Roberson--Walker metric, 

\begin{equation}
\label{metrica}
	ds^2=-dt^2+a(t)^2\delta_{ij}dx^idx^j, i,j=1,2,3,
\end{equation}

witch depicts a homogeneous and isotropic Universe described by the cosmic scale factor $a(t)$ which further depends on the cosmic time $t$. The present cosmological model takes into consideration a direct coupling between the Einstein tensor and the energy--momentum tensor, indicated by the following fundamental action \cite{Marciu:2023jvs}: 

\begin{equation}
\label{actiunee}
	S=S_m+\int d^4x \sqrt{-\tilde{g}} \big[f(R)+g(T^G) \big],
\end{equation}

with 

\begin{equation}
	T^G=G_{\mu \nu} T^{\mu \nu}.
\end{equation}

The energy--momentum tensor can be written according to the definition,

\begin{equation}
	T_{\mu \nu}=-\frac{2}{\sqrt{-\tilde{g}}}\frac{\delta(\sqrt{-\tilde{g}}L_m)}{\delta \tilde{g}^{\mu \nu}}.
\end{equation}

Next, in our calculations we shall use the variation of the energy--momentum tensor with respect to the inverse metric. For this, we shall consider the following relation \cite{Marciu:2023jvs},

\begin{equation}
\label{eq11}
	\frac{\delta T_{\alpha \beta}}{\delta \tilde{g}^{\mu \nu}}=\frac{\delta \tilde{g}_{\alpha \beta}}{\delta \tilde{g}^{\mu \nu}}L_m+\frac{1}{2}\tilde{g}_{\alpha \beta}L_m \tilde{g}_{\mu \nu}-\frac{1}{2}\tilde{g}_{\alpha\beta}T_{\mu\nu}-2\frac{\partial^2 L_m}{\partial \tilde{g}^{\mu\nu} \partial \tilde{g}^{\alpha\beta}}.
\end{equation}

\par 
Before proceeding to the presentation of the corresponding Friedmann equations we define the following relation, obtained by using the previous equation, eq.\eqref{eq11} \cite{Marciu:2023jvs}, 

\begin{equation}
	\Sigma_{\mu\nu}=G^{\alpha\beta}\frac{\delta T_{\alpha\beta}}{\delta \tilde{g}^{\mu\nu}}=-G_{\mu\nu}L_m+\frac{1}{2}G^{\alpha\beta}\tilde{g}_{\alpha\beta}(\tilde{g}_{\mu\nu}L_m-T_{\mu\nu})
	\\-2 G^{\alpha\beta}\frac{\delta^2 L_m}{\delta \tilde{g}^{\mu\nu} \delta \tilde{g}^{\alpha\beta}}.
\end{equation}

Note that in our action we have considered that a special decomposition is taken into account, by decomposing the full action into a term that is described by the scalar curvature ($f(R)$), and another component which describes the matter geometry interplay ($g(T^G)$).

\par 
Next, if we consider the variation of the term which is described by the scalar curvature in eq. \eqref{actiunee}, we obtain the following expression for the energy--momentum tensor \cite{Marciu:2023jvs}, 

\begin{equation}
	T_{\mu\nu}^{f(R)}=\tilde{g}_{\mu\nu}f(R)-2 R_{\mu\nu}f_R+2\nabla_{\mu}\nabla_{\nu}f_R-2 \tilde{g}_{\mu\nu} \Box f_R.
\end{equation} 

\par 
On the other hand, the variation of the matter--geometry interplay term in eq. \eqref{actiunee} leads to the corresponding energy momentum tensor \cite{Marciu:2023jvs}, 

\begin{multline}	T_{\mu\nu}^{g(T^G)}=\tilde{g}_{\mu\nu}g(T^G)+g_{,T^G}T R_{\mu\nu}-2 g_{,T^G} G_{\nu}^{\beta}T_{\mu\beta}-2 g_{,T^G} G_{\mu}^{\alpha}T_{\nu\alpha}
	\\-g_{,T^G}R T_{\mu\nu}-\Box (g_{,T^G} T_{\mu\nu})+\nabla_{\alpha}\nabla_{\mu}(g_{,T^G}T_{\nu}^{\alpha})+\nabla_{\alpha}\nabla_{\nu}(g_{,T^G}T_{\mu}^{\alpha})
	\\-\tilde{g_{\mu\nu}}\nabla_{\alpha}\nabla_{\beta}(g_{,T^G} T^{\alpha\beta})+\tilde{g_{\mu\nu}}\Box(g_{,T^G}T)-\nabla_{\mu}\nabla_{\nu}(g_{,T^G}T)-2g_{,T^G}\Sigma_{\mu\nu},
\end{multline}

where

\begin{equation}
	g_{,T^G}=\frac{d g(T^G)}{d T^G}.
\end{equation}

\par 
The variation of the action \eqref{actiunee} with respect to the inverse metric leads to the following modified Friedmann equations \cite{Marciu:2023jvs}, 

\begin{equation}
    \label{eqwd}T_{\mu\nu}^{f(R)}+T_{\mu\nu}^{g(T^G)}+T_{\mu\nu}^{m}=0.
\end{equation}

Taking the divergence of eq. \eqref{eqwd} we obtain the corresponding continuity equation,

\begin{equation}
	\nabla^{\mu}\big[T_{\mu\nu}^{f(R)}+T_{\mu\nu}^{g(T^G)}+T_{\mu\nu}^{m}\big]=0,
\end{equation}

describing various evolutionary aspects of our cosmological model. For the Roberson--Walker metric \eqref{metrica} we use the following definition of the matter stress--energy tensor, 

\begin{equation}
	T_{\mu\nu}=(\rho_m+p_m)u_{\mu}u_{\nu}+p_m g_{\mu\nu},
\end{equation} 

which is associated to a standard barotropic matter fluid. In this case, we obtain the following modified Friedmann equations \cite{Marciu:2023jvs},  

\begin{equation}
	\label{frconstraint}
	f(R)-6 f_R(\dot{H}+H^2)+6 H \dot{f_R}=\rho_m-g(T^G)-6 \rho_m g_{,T^G} \dot{H},
\end{equation}

\begin{equation}
	\label{acceleration}
	f(R)-2 f_R(\dot{H}+3 H^2)+2 \ddot{f_R}+4 H \dot{f_R}=-p_{T^G},
\end{equation}

 with 

\begin{equation}
		p_{T^G}=g(T^G)-2 g_{,T^G}(\rho_m(3 H^2+\dot{H})+H \dot{\rho_m}) -6 H^2 \rho_m \big[2 \rho_m \dot{H}+H \dot{\rho_m} \big] g_{,T^G T^G}.
\end{equation}

\par 
Finally, for our cosmological model we can define the effective (total) equation of state, 

\begin{equation}
	w_{tot}=-1-\frac{2}{3}\frac{\dot{H}}{H^2}.
\end{equation}
 
\par 
For the $f(R, G_{\mu\nu}T^{\mu\nu})$ gravity theory  \cite{Marciu:2023jvs} the phase space analysis showed the compatibility with the recent evolution of our Universe, explaining the late time cosmic acceleration as a physical effects due to the matter-geometry interplay. Note that in this section we have neglected the matter pressure, considering that $w_m=0$.

\subsection{The scalar tensor representation for $f(R, G_{\mu\nu}T^{\mu\nu})$ gravity}

\par 
In order to apply the formalism of irreversible thermodynamics for open systems, we need to obtain the scalar tensor representation for the $f(R, G_{\mu\nu}T^{\mu\nu})$ gravity theory, following Ref.~\cite{Cipriano:2023yhv} and references therein. Within the present paper, we shall investigate the cosmological model described by the following action, 

\begin{equation}
\label{actiune2}
    S=S_m+\frac{1}{2} \int d^4x \sqrt{-\tilde{g}} \big[f(R,T^G) \big],
\end{equation}

where $S_m$ describes the matter sector (as the dark matter component). The action for the $f(R, G_{\mu\nu}T^{\mu\nu})$ gravity theory can be written in an equivalent manner in the scalar tensor representation. To this regard, we need to introduce two scalar components $\alpha$ and $\beta$ which will further describe the pure geometrical context induced by the scalar curvature dependence and the matter--geometry interplay encapsulated into the $T^G=G_{\mu\nu}T^{\mu\nu}$ component. Hence, the equivalent action will be written in the following mode, 

\begin{equation}
\label{actrepr}
    S=\frac{1}{2} \int d^4x \sqrt{-\tilde{g}} \big[f(\alpha,\beta)+f_{\alpha}(R-\alpha)+f_{\beta}(T^G-\beta) \big].
\end{equation}

Since at this point the full action is described by two auxiliary scalar fields, we need to perform the variation of the previous action with respect to the scalar quantities. By performing the variation with respect to the scalar field $\alpha$ we get the following relation,

\begin{equation}
\label{eq22}
    f_{\alpha\alpha}(R-\alpha)+f_{\beta\alpha}(T^G-\beta)=0,
\end{equation}

while for the $\beta$ field we obtain:

\begin{equation}
\label{eq23}
    f_{\alpha\beta}(R-\alpha)+f_{\beta\beta}(T^G-\beta)=0.
\end{equation}

\par 
The system composed by the equations \eqref{eq22}--\eqref{eq23} can be written in the matrix representation as:
\begin{equation}
    M \cdot x = 0,
\end{equation}
with
\begin{equation}
    M=\begin{pmatrix}
f_{\alpha\alpha} & f_{\beta\alpha}\\
f_{\alpha\beta} & f_{\beta\beta}
\end{pmatrix} \cdot \begin{pmatrix}
R- \alpha\\
T^{G}-\beta
\end{pmatrix}=0 .
\end{equation}

\par 
The system of equations \eqref{eq22}--\eqref{eq23} has a unique solution if the following inequality is satisfied, 
\begin{equation}
    f_{\alpha\alpha} f_{\beta\beta} \neq f_{\alpha\beta}^2, 
\end{equation}
described by 
\begin{equation}
    \alpha=R, 
\end{equation}
and
\begin{equation}
    \beta=T^{G}.
\end{equation}

\par 
If we further introduce two scalar quantities as
\begin{equation}
    \phi=\frac{\partial f}{\partial \alpha}, 
\end{equation}

\begin{equation}
    \psi=\frac{\partial f}{\partial \beta}, 
\end{equation}

mediated by an auxiliary potential term

\begin{equation}
    V(\phi, \psi)=-f(\alpha, \beta)+\phi \alpha + \psi \beta, 
\end{equation}
we can rewrite the original action \eqref{actiune2} in the scalar tensor representation \eqref{actrepr}:
\begin{equation}
\label{actiunefinala}
    S=\frac{1}{2} \int d^4x \sqrt{-\tilde{g}} \big[\phi R+ \psi T^G - V(\phi, \psi) \big], 
\end{equation}

where

\begin{equation}
    V(\phi, \psi),\phi=R, 
\end{equation}

and 

\begin{equation}
    V(\phi, \psi), \psi=T^G.
\end{equation}

For the final action \eqref{actiunefinala} in the scalar tensor representation we shall perform the variation with respect to the inverse metric, obtaining the corresponding energy--momentum tensor, 

\begin{multline}
	T_{\mu\nu}=\phi G_{\mu\nu}+ g_{\mu\nu} \Box \phi-\nabla_{\mu}\nabla_{\nu}\phi+\frac{1}{2}g_{\mu\nu}V(\phi, \psi)+(-\frac{1}{2})g_{\mu\nu}\psi G^{\alpha\beta}T_{\alpha\beta}+\psi \Sigma_{\mu\nu}+\psi T_{\mu}^{\beta}G_{\nu\beta}+\psi T_{\nu\alpha}G_{\mu}^{\alpha}+\frac{1}{2}\psi T_{\mu\nu}R
 \\
 +(-\frac{1}{2})(\nabla_{\alpha}\nabla_{\mu}(\psi T_{\nu}^{\alpha}+\nabla_{\alpha}\nabla_{\nu}(\psi T_{\mu}^{\alpha})))+\frac{1}{2}\Box (\psi T_{\mu\nu})+\frac{1}{2} g_{\mu\nu} \nabla_{\alpha} \nabla_{\beta} (\psi T^{\alpha\beta})-\frac{1}{2}\psi T R_{\mu\nu}+\frac{1}{2}\nabla_{\mu}\nabla_{\nu}(\psi T)-\frac{1}{2}\Box (\psi T) g_{\mu\nu}.
\end{multline}

\par 
In this case the corresponding modified Friedmann relations are the following:

\begin{equation}
    3 H^2 \phi=\rho_{m}+\frac{1}{2}  V(\phi, \psi)-\frac{3}{2}\psi \rho_{m}H^2-3 \psi \rho_{m}\dot{H}-3 H \dot{\phi},
\end{equation}

\begin{equation}
     V(\phi, \psi)+H^2(-6\phi+3\rho_{m}\psi)+2 H (\psi \dot{\rho_{m}}-2 \dot{\phi}+\rho_{m}\dot{\psi})
     \\
     -2(2 \phi \dot{H}-\rho_{m} \psi \dot{H}+\ddot{\phi})=0,
\end{equation}

specific for the Robertson--Walker metric \eqref{metrica} in the scalar tensor representation. We note that these equations contain second order terms, describing a cosmological model where the matter--geometry interplay is mediated by the evolution of two additional scalar fields $\phi$ and $\psi$. As in the previous case, the matter component is assumed to be pressure--less, with $w_m=0$.

\section{Overview of thermodynamics in open systems}
\label{termo}

\par 
In what follows, we shall present briefly for completeness various theoretical aspects related to the thermodynamics of open systems, following mainly Ref.~\cite{Cipriano:2023yhv}. In the case of a particular adiabatic process, we can write the energy conservation relation as follows,

\begin{equation}
\label{reltermo}
    d(\rho V)+p dV=\frac{h}{n} d(n V),
\end{equation}
valid for a specific open system described by a number of particles $N$ contained in a volume $V$. In this relation we have denoted the particle number density with $n=N/V$, while $h=\rho+p$ represents the enthalpy per volume (in units). As usual, $\rho$  represents the energy density, and $p$ the associated pressure. Taking into account that we have an open system, we can write the second law of thermodynamics, by encapsulating the entropy flow $d_{e}S$ and the entropy creation component $d_{i}S$, 

\begin{equation}
    dS=d_{e}S+d_{i}S \geq 0.
\end{equation}

This inequality is associated to the evolution of the total entropy, showing the specific growth. Next, for the entropy flow we can add the following relation  \cite{Pinto:2022tlu}

\begin{equation}
    d_{e}S=\frac{dQ}{T},
\end{equation}

while for the entropy creation term we have \cite{Pinto:2022tlu}, 

\begin{equation}
    d_{i}S=\frac{s}{n} d(n V).
\end{equation}

Taking into account that the heat exchange is zero ($dQ=0$), we can write the previous second law of thermodynamics

\begin{equation}
    dS \geq 0.
\end{equation}

This specific expression shows that the variation of the entropy should be positive, determined by the creation of new particles, assuming that the system is adiabatic. This inequality implies that, assuming a thermodynamic interpretation, the creation of matter due to the matter--geometry interplay is an irreversible process.    

\par 
Next, we shall assume that we have a homogeneous and isotropic cosmological system described by the Robertson--Walker metric \eqref{metrica}, contained in a comoving volume $V=a^3$. Taking into account these considerations, the energy conservation relation reduces to

\begin{equation}
    \frac{d}{dt} (\rho a^3)+ p \frac{d}{dt} a^3=\frac{\rho + p}{n} \frac{d}{dt} (n a^3).
\end{equation}

If we introduce the definition of Hubble parameter, we obtain a continuity--like equation, taking into consideration a thermodynamic interpretation approach, 

\begin{equation}
\label{exprrr}
    \dot{\rho} + 3 H (\rho + p)=\frac{\rho+p}{n}(\dot{n}+3 H n).
\end{equation}

Furthermore, we can write the number current for the cosmological fluid, 

\begin{equation}
    N^{\mu}=n u^{\mu}.
\end{equation}

Taking the corresponding divergence in the comoving frame we obtain

\begin{equation}
    \nabla_{\mu} N^{\mu}= \dot{n}+3Hn=n\Gamma,
\end{equation}

reducing the previous expression \eqref{exprrr} to the following form:

\begin{equation}
    \dot{\rho} + 3 H (\rho + p)=(\rho+p)\Gamma.
\end{equation}

From the previous expression we deduce the expression of the matter/particle creation rate $\Gamma$, 

\begin{equation}
    \Gamma=\frac{\dot{\rho} + 3 H (\rho + p)}{\rho+p}.
\end{equation}

\par 
In the case of open systems, an auxiliary pressure can be introduced ($p_{c}$ as the creation pressure), expressing the thermodynamic energy conservation relation \eqref{reltermo} in an alternative manner,  

\begin{equation}
    \frac{d}{dt}(\rho a^3)+(p+p_{c})\frac{d}{dt}a^3=0.
\end{equation}

Finally, at this point we obtain the definition for the creation pressure introduced earlier,  

\begin{equation}
    p_{c}=-\frac{\rho+p}{3 H} \Gamma,
\end{equation}

observing the specific dependence on the particle creation rate, the energy density and pressure for the cosmological fluid, encapsulating physical effects due to the cosmic expansion through the Hubble parameter.   
\par 
As shown in \cite{Cipriano:2023yhv}, the time evolution of the entropy depends on the creation rate, 

\begin{equation}
    S(t)=S_{0} e^{ \int_{0}^{t} \Gamma(t^{'}) \,dt^{'}},
\end{equation}

while the temperature, defined as follows, 

\begin{equation}
    T(t)=T_{0} e^{c_{s}^2 \int_{0}^{t'} (\Gamma(t')-3 H) \,dt'},
\end{equation}

is influenced by the matter creation rate, the Hubble parameter, and the associated speed of sound, $c_s=\sqrt{{(\partial p/\partial n)}_n}$.

\section{Analytical and numerical properties}
\label{analiza}
\par 
In this section, we shall discuss various analytical and numerical properties for the present cosmological model which takes into consideration a matter--geometry interplay.  As previously mentioned, in our analysis we have considered that the background dynamics corresponds to a flat scenario where the cosmological fluid behaves as a pressure--less fluid with $w_m=0$.
\par 
As a first step, we have verified that our model incorporates the de Sitter solution. For this solution in the asymptotic regime $H=H_0=$constant, describing the late time acceleration epoch of our Universe. Taking into consideration the vacuum case where $\rho=p=0$ and $H=H_0=$constant, we have obtained the same solution as in Ref.~\cite{Cipriano:2023yhv}, without a particle creation rate ($\Gamma=0$),  showing that the matter creation might appear in different stages in the evolution of the Universe in the non--vacuum case. For further details on this solution, see Ref.~\cite{Cipriano:2023yhv} and references therein.

\par 

Furthermore, as a second step we have considered the case of the constant energy density, which implies $\rho=\rho_0$, $p=0$, $H=H_0=$constant. Following Ref.~\cite{Cipriano:2023yhv}, we have obtained similar results, having the creation rate $\Gamma=3 H_0$, with the creation pressure $p_c=-\rho_0$. This is somehow expected since at the fundamental level the coupling between the Einstein tensor and the energy--momentum tensor implies a direct connection between the evolution of the Hubble parameter and the growth of the energy density.

\subsection{The case where the energy density is time--dependent}

\par 
In this section we shall discuss the asymptotic case where the Hubble expansion rate is constant, ($H_0=$const.), while the matter energy density is a time-depending variable. As in the previous discussion, the matter is assumed to be a pressure--less fluid. The first equation of the potential reduces to 

\begin{equation}
    \frac{dV(\phi, \psi)}{d\phi}=R=6(\dot{H}+2H^2)\approx 12 H_0^2,
\end{equation}

implying that the potential should be proportional to $\phi$, $V(\phi, \psi) \propto 12H_0^2 \phi$. For the second argument of the potential we shall take into consideration a squared variation. Hence, the potential analyzed has the following dependence, 

\begin{equation}
    V(\phi, \psi)=\Lambda_0+12 H_0^2 \phi+\beta \psi^2.
\end{equation}
\par
For simplicity, we have considered that the constant term in the above relation is neglected, $\Lambda_0=0$. Here $\beta$ is a constant parameter which describes the strength of $\psi$ squared coupling term. The second equation of the potential reduces to,

\begin{equation}
    \frac{dV(\phi, \psi)}{d\psi}=3 H_0^2 \rho,
\end{equation}

implying a non--negligible relation between the variation of the matter energy density $\rho$ and the $\psi$ term, $3 H_0^2 \rho=2 \beta \psi$. Hence, in our asymptotic calculations we can replace the matter energy density with $\rho=\frac{2 \beta \psi}{3 H_0^2}$. The Friedmann constraint equation reduces in the asymptotic regime to 

\begin{equation}
    \frac{\beta  \psi ^2}{2}+3 H_0 \dot{\phi}-3 H_0^2 \phi =\frac{2 \beta  \psi }{3 H_0^2}.
\end{equation}

From this, we can obtain the algebraic solution for the $\psi$ variable, 

\begin{equation}
    \psi=\frac{2}{3 H_0^2} \pm \frac{\sqrt{2} \sqrt{2 \beta ^2-27 \beta  H_0^5 \dot{\phi}+27 \beta  H_0^6 \phi }}{3 \beta  H_0^2}, 
\end{equation}

restricted to the real domain. Next, for simplicity we take $H_0=1$ and $\beta=1$, presenting the solution in units of $H_0$. In our analysis we have observed that the first solution leads to viable physical effects, $\psi=\frac{2}{3}+\frac{1}{3} \sqrt{-54 \dot{\phi} +54 \phi +4}$. With this solution, we use the second Friedmann equation, the acceleration relation, and replace $\psi$ and its derivative, obtaining the final representation in the asymptotic regime, 

\begin{equation}
\label{eqnnum}
    2 \sqrt{-54 \dot{\phi}+54 \phi +4}+\frac{12 \sqrt{2} \dot{\phi}}{\sqrt{-27 \dot{\phi}+27 \phi +2}}-\frac{12 \sqrt{2} \ddot{\phi}}{\sqrt{-27 \dot{\phi}+27 \phi +2}}+36 \phi +4=15 \ddot{\phi}+21 \dot{\phi},
\end{equation}
 a differential equation for $\phi$ which can be solved numerically for different initial conditions. Using similar arguments, we can obtain the expression for the matter creation rate, 

 \begin{equation}
     \Gamma=\frac{\sqrt{-54 \dot{\phi}+54 \phi +4}-135 \dot{\phi}+135 \phi +2}{-75 \dot{\phi}+75 \phi +2},
 \end{equation}

 and the creation pressure, 

 \begin{equation}
 p_c=-\frac{2 \left(4 \sqrt{-54 \phi '+54 \phi +4}+27 \phi  \left(5 \sqrt{-54 \phi '+54 \phi +4}+12\right)-27 \left(5 \sqrt{-54 \phi '+54 \phi +4}+12\right) \phi '+8\right)}{27 \left(-75 \phi '+75 \phi +2\right)}.
 \end{equation}
\par
\begin{figure}[t]
  \includegraphics[width=8cm]{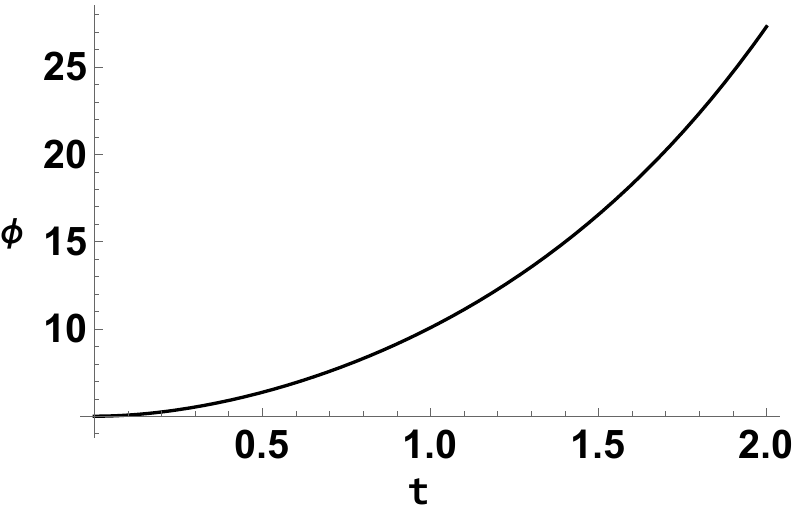} 
\caption{The evolution of the $\phi$ variable in the asymptotic regime where the Hubble expansion is saturated.}
\label{fig:as1}       
\end{figure}

\begin{figure}[t]
  \includegraphics[width=8cm]{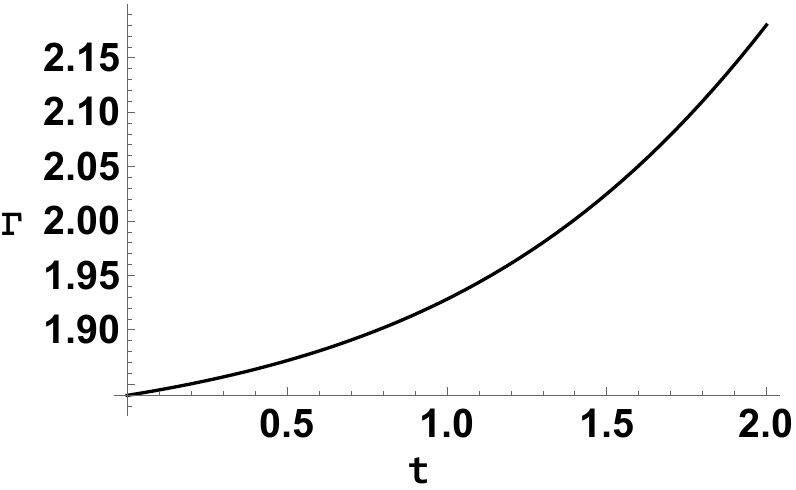} 
\caption{The evolution of the matter creation rate in the asymptotic case where the Hubble expansion is saturated.}
\label{fig:as2}       
\end{figure}

\begin{figure}[t]
  \includegraphics[width=8cm]{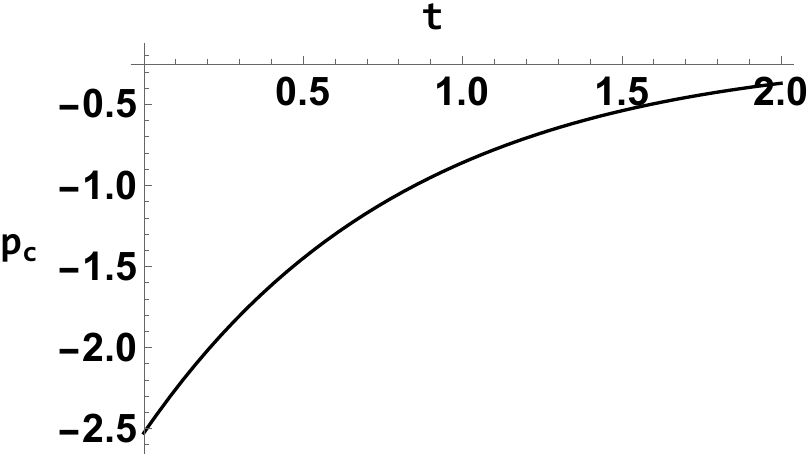} 
\caption{The evolution of the creation pressure in the asymptotic regime where the Hubble expansion is saturated.}
\label{fig:as3}       
\end{figure}

 
At this point, we numerically integrate the equation \eqref{eqnnum}, displaying the results in Figs.~\ref{fig:as1}--\ref{fig:as3}. The initial conditions affect the numerical values of the corresponding solutions, leading to distinct physical effects. The evolution of the $\phi$ variable is displayed in Fig.~\ref{fig:as1} for the following initial conditions, $\phi(0)=5, \dot{\phi(0)}=0.001$. In this case, the initial values have been fine tuned in order to obtain viable physical effects. Next, in Fig.~\ref{fig:as2} we depicted the matter creation rate and the evolution of the creation pressure in Fig.~\ref{fig:as3}. These graphs are viable also from a thermodynamic point of view, showing that the matter creation starts in the early stages of evolution, leading to a non--negligible creation pressure. Note that for this solution the creation rate is positive, satisfying the basic thermodynamic assumptions considered in the previous section.   

\subsection{Cosmological scenarios with a redshift dependence for the Hubble parameter}
\par 
In this section we shall present several approximate numerical solutions by encapsulating a specific parametrization for the Hubble characteristic expansion rate. This enables us to impose a specific evolution for the Hubble parameter, viable from an observational point of view. Hence, in our study we shall first obtain viable best fitted values in the case where a specific parametrization for the Hubble characteristic expansion rate is considered. To this regard, we shall take into consideration the following expansion rate \cite{Mamon:2016wow}, 

\begin{equation}
    \label{unu}
    H(z)=H_0 \sqrt{\Omega_{m0}(1+z)^3+(1-\Omega_{m0})(1+z)^{\alpha}e^{\beta z}}. 
\end{equation}

\par
\begin{figure}[t]
  \includegraphics[width=8cm]{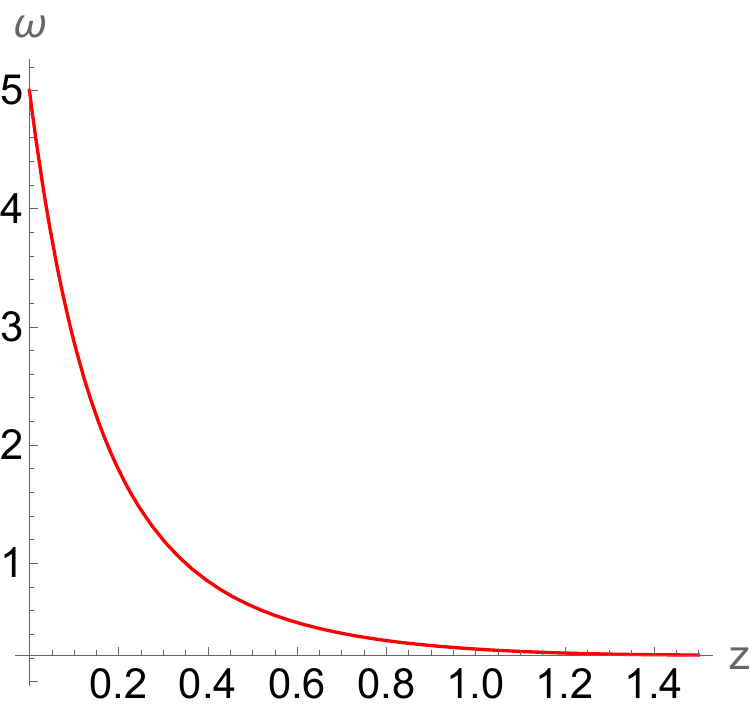} 
\caption{The evolution of the $\omega$ variable in the case of CC+BAO observations.}
\label{fig:b5}       
\end{figure}

 
This parametrization has been probed by Mamon et al. \cite{Mamon:2016wow} in the case of quintessence dark energy model minimally coupled with gravity. The analysis has been performed for Union 2.1 compilation and BAO/CMB data. The present study applies the previous mentioned parametrization using cosmic chronometers, baryon acoustic observations, presenting the confidence intervals for the corresponding parameters. For further details on the numerical implementation, see Ref.~\cite{Marciu:2023hdb} and references therein. The results of our analysis are presented in Fig.~\ref{fig:b5}, showing the best fitted values of the relevant cosmological parameters, and the specific one sigma intervals.  
\par 
After obtaining the best fitted values of the corresponding variables encoded into the Hubble expansion rate, we further encapsulate this solution into the field equations in the scalar tensor representation, obtaining the evolution of the matter creation pressure $p_c$ and the creation rate $\Gamma$ for this model. As the first step in the analysis we have considered the introduction of the following auxiliary variables, 

\begin{equation}
    H=H_0 h,
\end{equation}

\begin{equation}
    \tau=H_0 t,
\end{equation}

\begin{equation}
    \rho=3 H_0^2 r,
\end{equation}

\begin{equation}
    V=3 H_0^2 U,
\end{equation}

\begin{equation}
    \psi=\frac{\bar{\psi}}{9 H_0^2}.
\end{equation}

\par 
After changing the time dependence to the $\tau$ variable, we obtain the final relation for the first Friedmann equation, 

\begin{equation}
    -\frac{1}{2}U(\phi, \bar{\psi})+h^2 \psi+\frac{r h^2 \bar{\psi}}{6}+\frac{1}{3} r \bar{\psi} \frac{dh}{d \tau}+ h \frac{d \phi}{d \tau}=r.
\end{equation}

\par 
In this case the potential equation reduces to, 

\begin{equation}
    \frac{dU(\phi, \bar{\psi})}{d \phi}=2 (\frac{dh}{d \tau}+2 h^2),
\end{equation}

\begin{equation}
    \frac{dU(\phi, \bar{\psi})}{d \bar{\psi}}=\frac{1}{3} r h^2.
\end{equation}

\par 
The acceleration equation reduces to the following expression, 

\begin{equation}
    2 \phi \frac{dh}{d \tau}+\frac{d^2 \phi}{d \tau^2}=h(\frac{1}{3}\bar{\psi} \frac{dr}{d \tau}+\frac{d \phi}{d \tau})+ 3r(-1+\frac{1}{3}h^2 \bar{\psi}+\frac{4}{9} \bar{\psi} \frac{dh}{d \tau}+\frac{1}{9} h \frac{d \bar{\psi}}{d \tau}).
\end{equation}

\par 
Moreover, for simplicity we introduce an additional variable, 

\begin{equation}
    \omega=\frac{d \phi}{d \tau}.
\end{equation}

\par 
Next, we change the $\tau$ dependence of the dynamical system to the redshift variable $z$, considering the time dependence in units of $H_0$ ($H_0=1$). In this case, we use the following relations, 

\begin{equation}
    1+z=\frac{1}{a},
\end{equation}

\begin{equation}
    \frac{d}{d \tau}=-(1+z) h \frac{d}{dz}.
\end{equation}

\par 
Taking into account the previous mentioned considerations, we obtain the final expressions which describe the cosmological system, 

\begin{equation}
    \omega=-(1+z) h \frac{d \phi}{dz},
\end{equation}

\begin{equation}
\label{eqax}
    \frac{dU}{d \psi}=\frac{1}{3} r h^2,
\end{equation}

\begin{equation}
    \frac{dU}{d \psi}=2(2h^2-(1+z)h \frac{dh}{dz}),
\end{equation}

\begin{equation}
    -\frac{1}{2}U(\phi, \psi)+h^2 \phi+\frac{r h^2 \psi}{6}-\frac{1}{3}r \psi(1+z)h \frac{dh}{dz}+h \omega=r, 
\end{equation}

\begin{equation}
\label{eqay}
    -2 \phi (1+z) h \frac{dh}{dz}-(1+z)h \frac{d \omega}{dz}=h(-\frac{1}{3}\psi (1+z)h \frac{dr}{dz}+\omega)+3r(-1+\frac{1}{3}h^2 \omega-\frac{4}{9}\psi(1+z)h \frac{dh}{dz}-\frac{1}{9}h^2(1+z)\frac{d \psi}{dz}).
\end{equation}

\begin{figure}[t]
  \includegraphics[width=8cm]{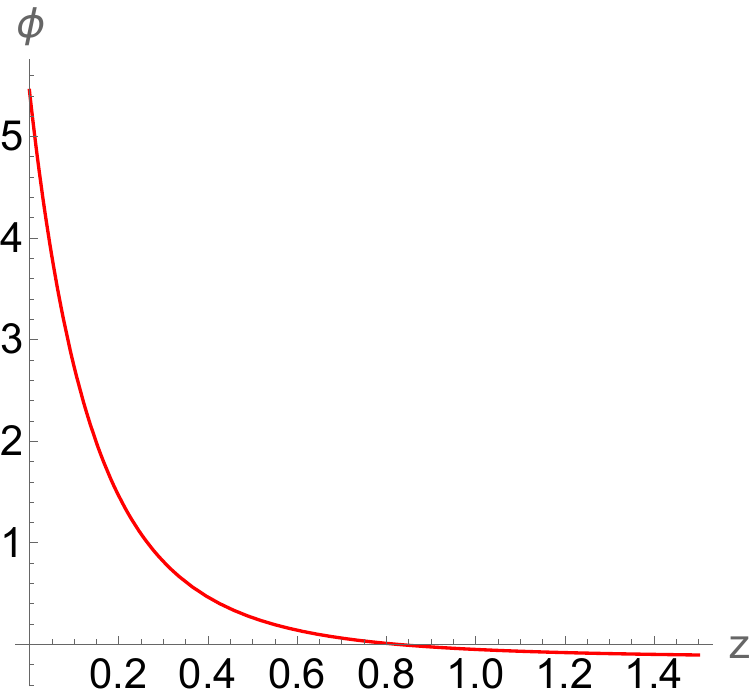} 
\caption{The redshift dependence of the $\phi$ variable in the asymptotic case which takes into account CC+BAO observations.}
\label{fig:b6}       
\end{figure}

\begin{figure}[t]
  \includegraphics[width=8cm]{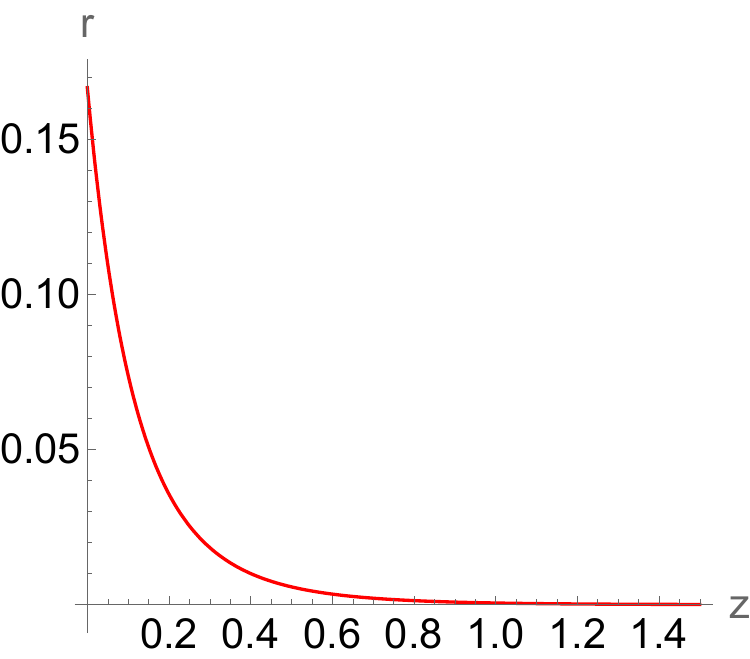} 
\caption{The dynamics for the $r$ variable in the asymptotic case which takes into account CC+BAO observations.}
\label{fig:b7}       
\end{figure}

\begin{figure}[t]
  \includegraphics[width=8cm]{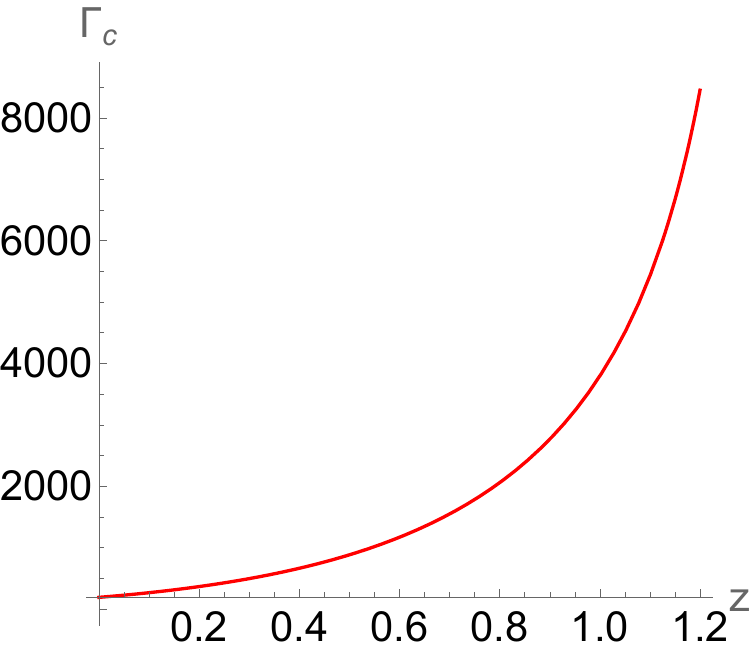} 
\caption{The evolution of the matter creation rate in the asymptotic case where the Hubble expansion is constrained using CC+BAO observations.}
\label{fig:b8}       
\end{figure}

\begin{figure}[t]
  \includegraphics[width=8cm]{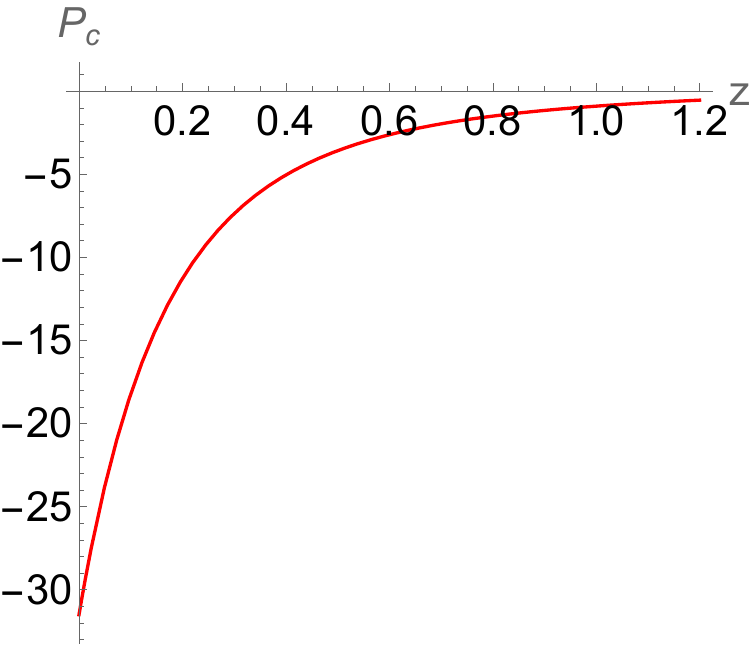} 
\caption{The evolution of the creation pressure in the asymptotic case where the Hubble expansion is constrained using CC+BAO observations.}
\label{fig:b9}       
\end{figure}

Note that we have dropped out the over line for the $\psi$ variable, for simplicity. In our analysis we have considered that the potential is decomposed as follows, $U(\phi, \psi)=\tilde{\alpha} \phi^n+\tilde{\beta} \psi^m+\tilde{\gamma} \phi \psi$, with $\tilde{\alpha}, \tilde{\beta}, \tilde{\gamma}, n, m$ constant parameters. In the analysis we have integrated this system of equations in redshift, considering that the $h$ variable is described by the previous mentioned representation in eq. \eqref{unu}. The results of the analysis are presented in Figs.~\ref{fig:b5}--\ref{fig:b9} for the following specific values: $\tilde{\alpha}=-0.001, \tilde{\beta}=+0.001, \tilde{\gamma}=0.01, \omega(z=0)=5, m=1, n=1$. The evolution of the $\omega$ variable in the asymptotic case which takes into account CC+BAO observations is presented in Fig.~\ref{fig:b5}. For the $\phi$ variable the evolution is shown in Fig.~\ref{fig:b6}, while for $r$ quantity the dependence is depicted in Fig.~\ref{fig:b7}. The redshift evolution of most important thermodynamic quantities are shown in  Figs.~\ref{fig:b8}--\ref{fig:b9}. In Fig.~\ref{fig:b8} we have represented the redshift dependence of the matter creation rate $\Gamma$. The evolution of the corresponding creation pressure $p_c$ is shown in Fig.~\ref{fig:b9} for specific initial conditions. We can observe that the matter creation rate $\Gamma$ is positive and increases towards a maximum in the early stages of evolution. For the creation pressure $p_c$ we have a similar evolution, at late times ($z \to 0$) the creation pressure is negative implying a matter creation mechanism, while in the early times ($z \to 1.2 $) the matter creation evolves asymptotically towards zero. Note that the numerical solution obtained by fine--tuning is specific for a matter creation mechanism, satisfying the fundamental thermodynamic assumption rules in the early stages of evolution.

\begin{figure}[t]
  \includegraphics[width=10cm]{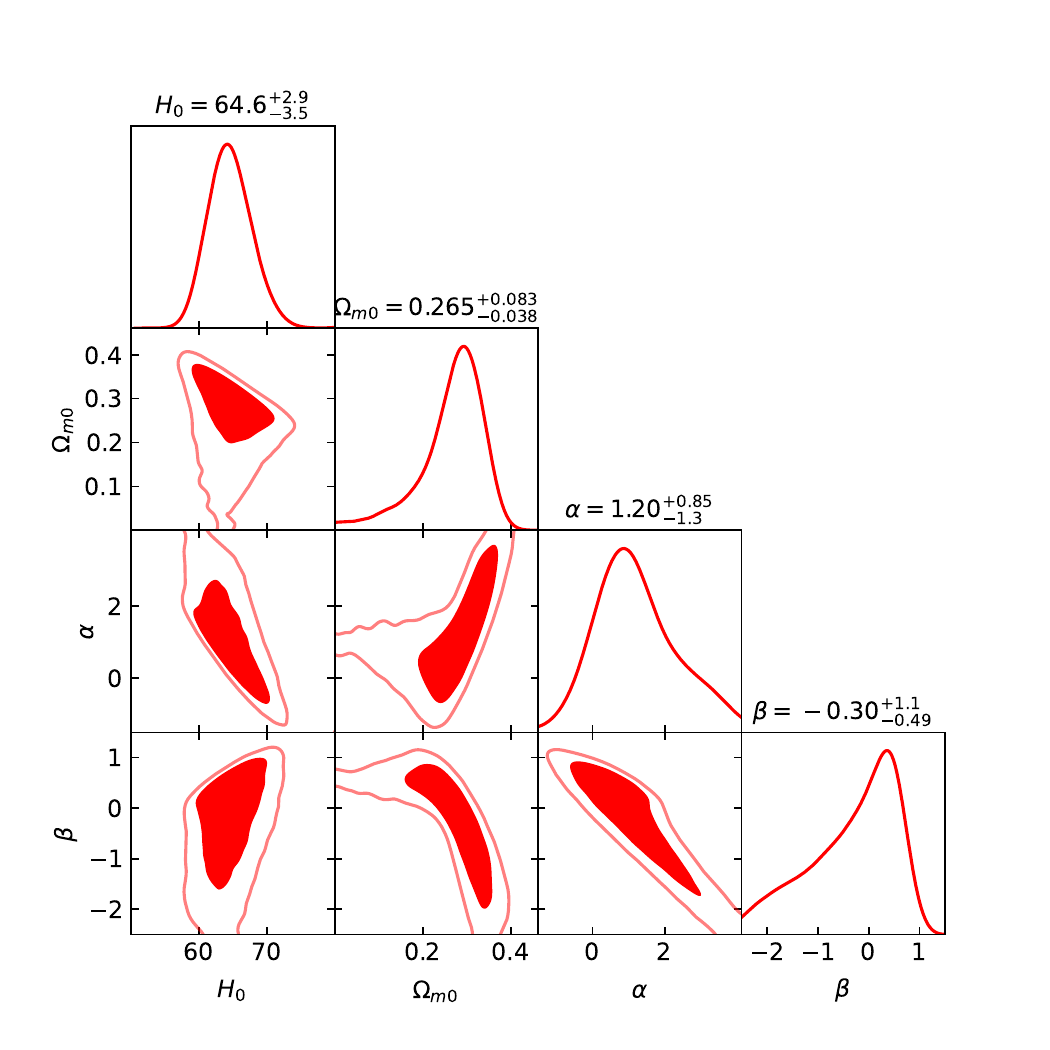} 
\caption{The posterior distributions in the case of CC+BAO observations.}
\label{fig:b1}       
\end{figure}

\begin{figure}[t]
  \includegraphics[width=10cm]{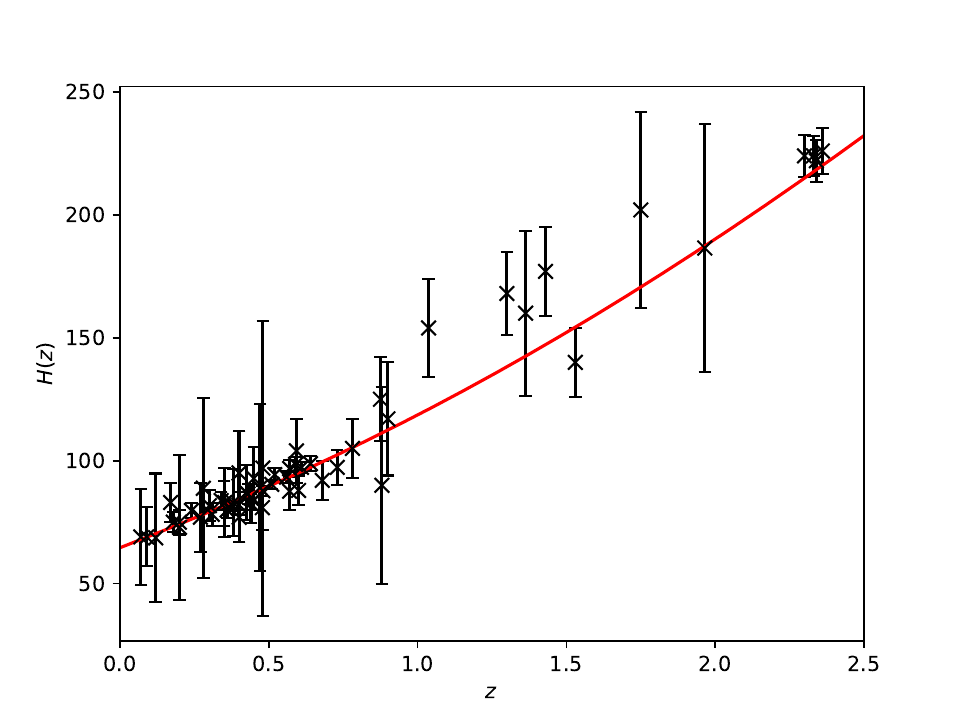} 
\caption{The evolution of the Hubble parameter in the case of CC+BAO observations. For the CC+BAO data we have used Ref.~\cite{Bernardo:2021ynf}.}
\label{fig:b2}       
\end{figure}

\subsection{Numerical evaluation of the field equations in redshift}

\par
\begin{figure}[t]
  \includegraphics[width=8cm]{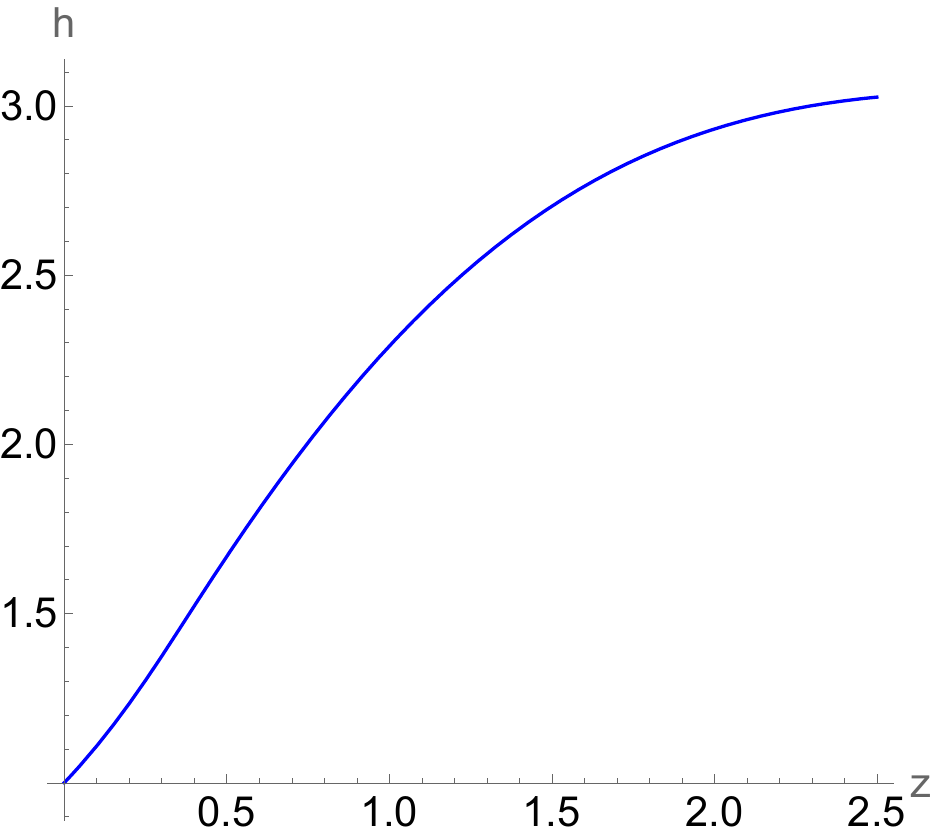} 
  \includegraphics[width=8cm]{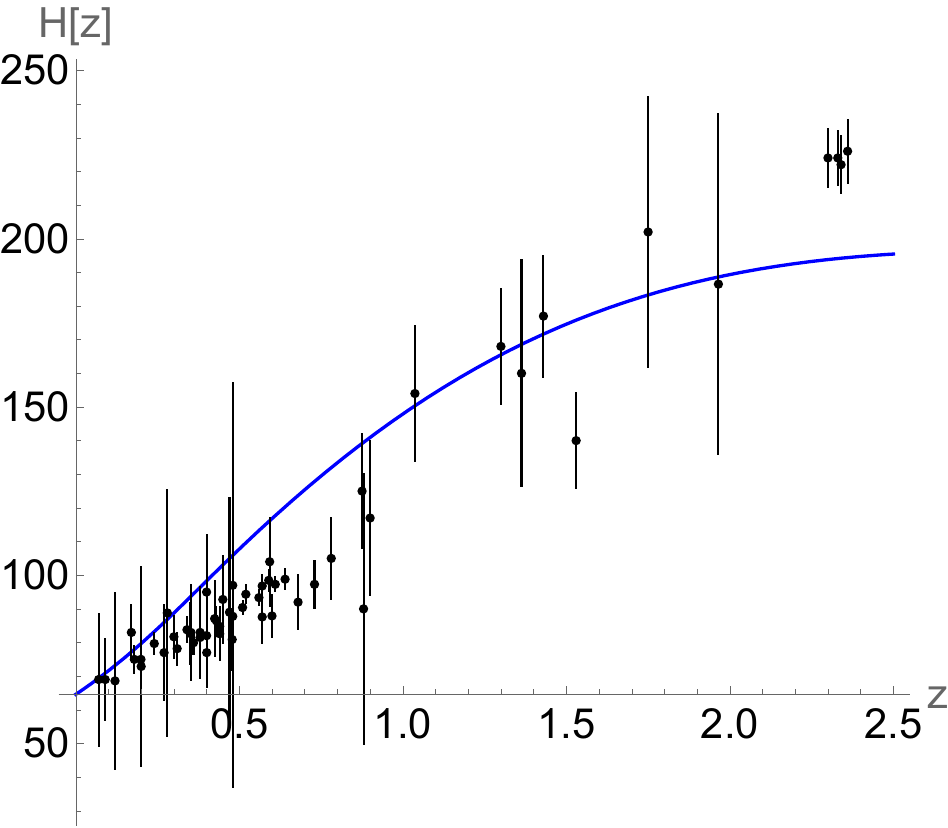} 
\caption{The evolution of the Hubble parameters in the case where the Friedmann equations are solved directly. We have considered that the Hubble rate in the present is: $H_0=64.6$, $H(z)=h(z) H_0$. The CC+BAO data is taken from \cite{Bernardo:2021ynf}.}
\label{fig:sol_has}       
\end{figure}

\begin{figure}[t]
  \includegraphics[width=8cm]{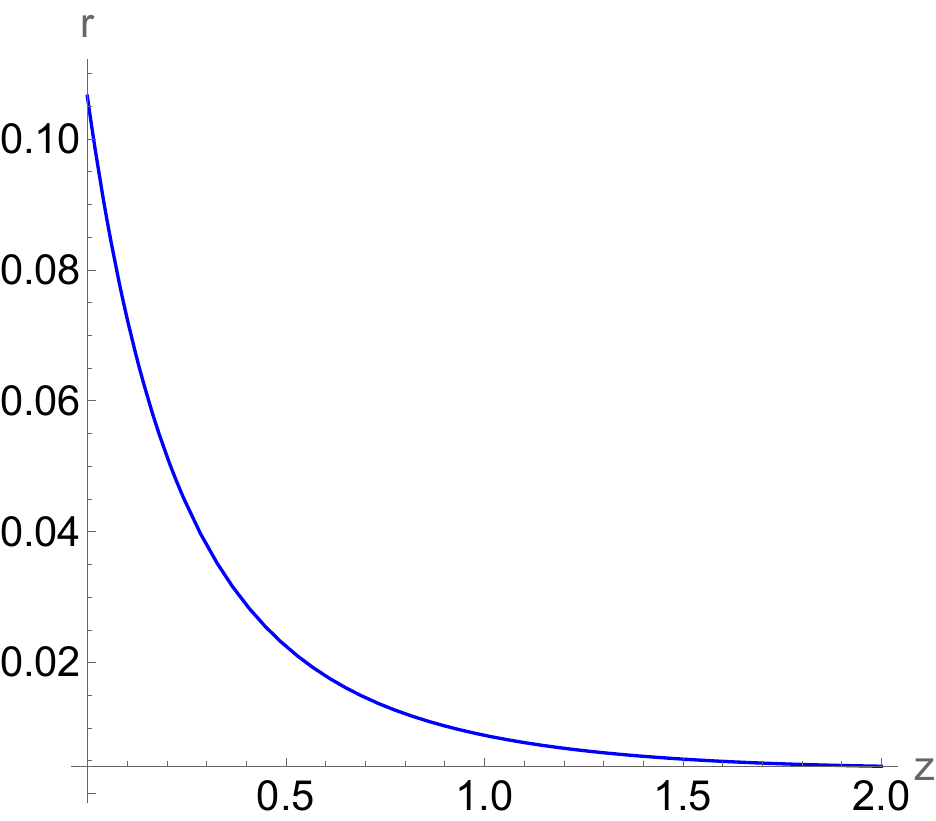}  
\caption{The dynamics for the $r$ variable obtained by the direct integration of the field equations.}
\label{fig:sol_r}       
\end{figure}

\begin{figure}[t]
  \includegraphics[width=8cm]{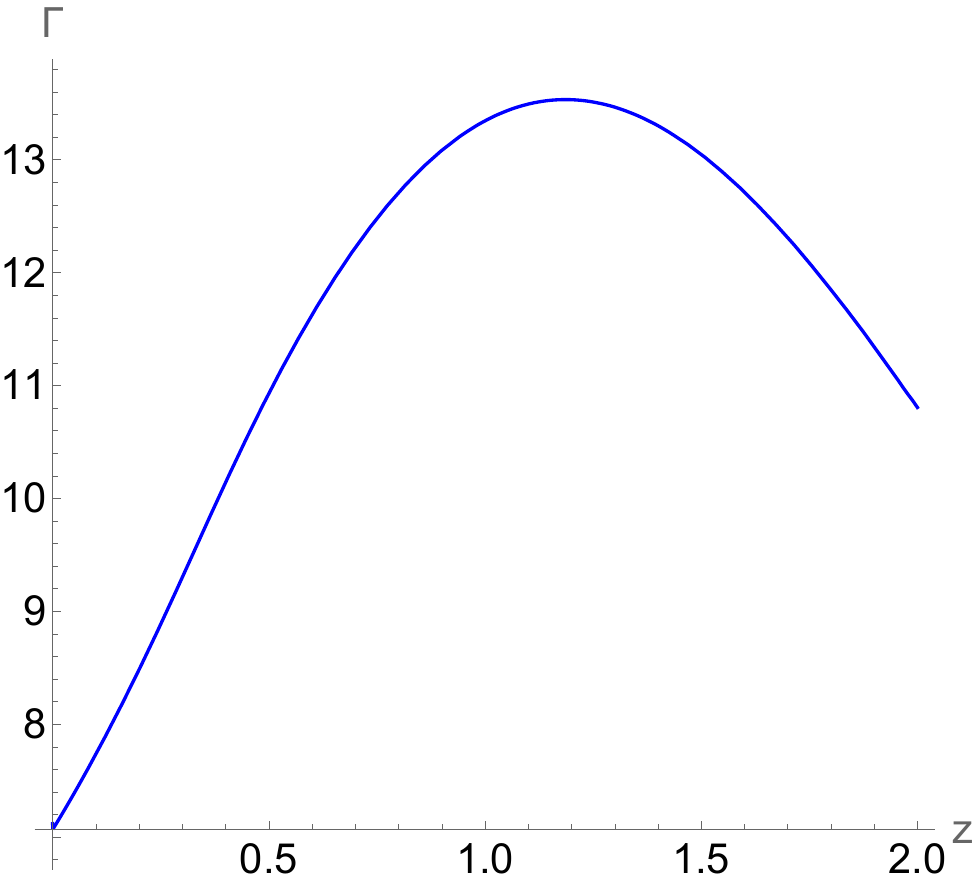} 
  \includegraphics[width=8cm]{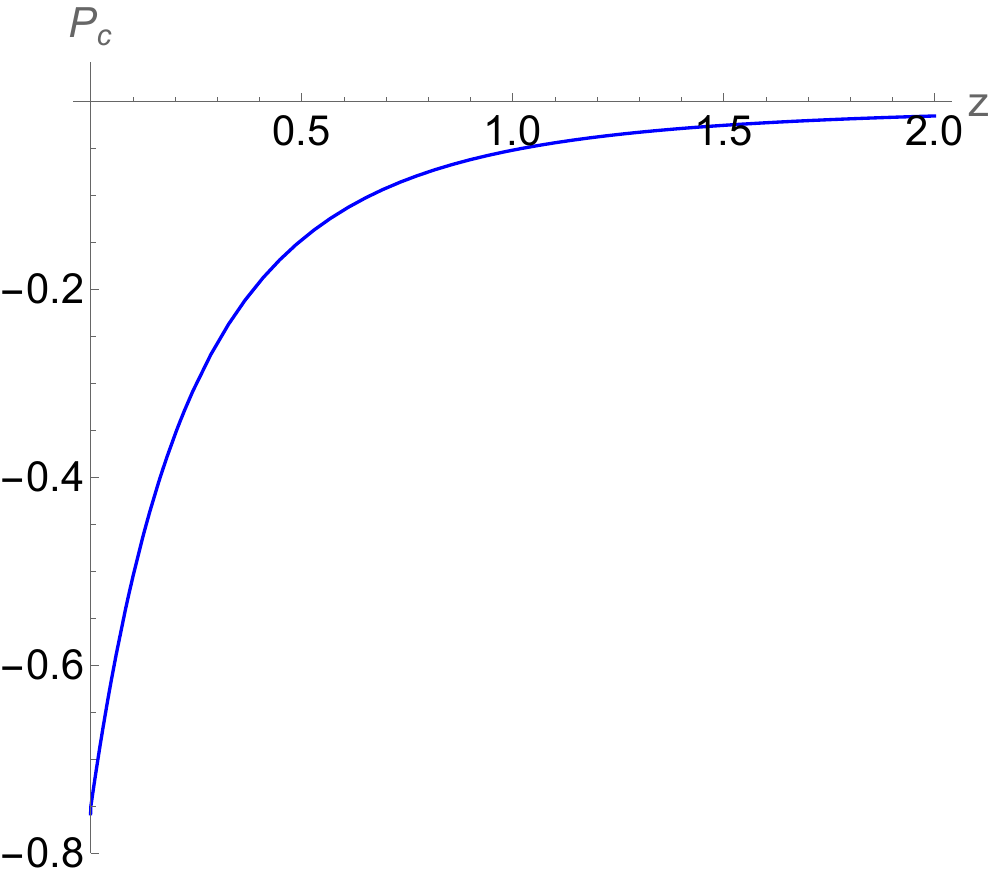} 
\caption{The matter creation rate $\Gamma$ and the creation pressure $p_c$ obtained by direct integration.}
\label{fig:sol_Gamma_presiune}       
\end{figure}

 
In what follows, we shall also consider the case where the field equations \eqref{eqax}--\eqref{eqay} are directly integrated, without any parametrization for the Hubble expansion rate. The results of the direct integration are presented in Figs.~\ref{fig:sol_has}--\ref{fig:sol_Gamma_presiune}. For the integration we have considered the following values:  $\tilde{\alpha}=0, \tilde{\beta}=0, \tilde{\gamma}=0.355, m=1, n=1, h(z=0)=1, h'(z=0)=1, \phi(z=0)=0.1, \phi'(z=0)=-0.2$. In this case, we can observe that the evolution corresponds to a non--negligible matter creation with a significant creation pressure in the late times. From the graphs, we can note that for the matter creation rate we see a dynamics corresponding to a significant bounce in the near past. The dimensionless matter density $r$ evolves asymptotically, having a significant value in the present days. Furthermore, the evolution of the Hubble parameter is comparable to the observed values, presenting a minor tension at high redshift. As expected, the results are satisfying the fundamental principles associated to the thermodynamic assumptions previously considered. Lastly, we mention here that the present results highly depend on the choice of the corresponding parameters, affecting the evolution of the field equations. 

\section{Conclusions}
\label{conclusions}

\par 
In this paper we have studied a novel dark energy model which takes into account a direct coupling between the Einstein tensor and the energy--momentum tensor. To this regard, we have analyzed the matter creation or annihilation in a specific model which encodes the interplay between matter and geometry at the fundamental level. After a brief introduction into the accelerated expansion phenomenon, we have presented the full action and the corresponding field equations which governs the specific dynamics. Then, we have deduced the scalar tensor representation in the case where the Einstein tensor is directly coupled with the energy--momentum tensor. Furthermore, we have presented several aspects related to the thermodynamics of open systems. In the following, we have obtained the final field equations in the case of the Robertson--Walker metric, deducing the relations for the creation (annihilation) pressure and the matter creation (annihilation) rate. In this manuscript, we have analyzed different specific solutions. In the first case,  we have checked the vacuum solution and the constant energy density solution in the analytical case. Then, we have deduced the numerical solution corresponding to the time--depending energy density when the Hubble expansion rate is saturated (the de--Sitter expansion), presenting various physical aspects. In the last part of this section, we have used the Markov chain Monte Carlo (MCMC) algorithms in the case of cosmic chronometers and baryon acoustic oscillations. In this case, we have used a specific parametrization for the Hubble parameter, obtaining specific posterior distributions for the corresponding cosmological parameters. The best--fitted parametrizations are then injected into the scalar tensor field equations, obtaining numerical solutions of the cosmological model by fine--tuning. For the numerical solutions obtained the model presents a non--zero creation rate in the near past and a corresponding creation pressure, an interesting result showing different aspects of the matter geometry interplay. Hence, the matter creation in a specific model which takes into account a possible interplay between matter and geometry is directly compatible to the accelerated expansion and the late time evolution of the Universe, affecting the background dynamics.

\section{data availability}
\label{sec:6}
Data sharing not applicable to this article as no datasets were generated or analysed during the current study.

\section{Acknowledgements}
The work was partially supported by the project 41PFE/30.12.2021, financed by the Ministry of Research, Innovation and Digitalization through Program 1 – Development of the National R$\and$D System, Subprogram 1.2. Institutional performance – Financing projects for excellence in RDI. For this manuscript we have used mainly Wolfram Mathematica \cite{Mathematica} and xAct \cite{xact}. The MCMC algorithms were implemented on the computer cluster using Python, Numpy, and Cobaya packages.  

\bibliography{sorsamp}

\end{document}